# Design and testing of an affordable desktop wind tunnel

Miguel de la Cruz[1] and Paolo Luzzatto-Fegiz[1]

[1]Department of Mechanical Engineering, University of California, Santa Barbara, CA 93106, USA

**Abstract:** Wind tunnels are a key source of data collection, but their cost and size can be a significant obstacle to their acquisition and usage, especially for applications such as instrument calibration, instruction, or in-class demonstrations. Here we propose a design for a cost-effective, desktop wind tunnel. This design takes advantage of readily available, inexpensive materials. Special consideration was taken to allow the wind tunnel to be serviceable, as well as giving the operator the ability to change key features without a complete redesign. There are three main sections, the first being a fan enclosure, which holds seven ducted fans in a hexagonal array. The second section holds honeycomb flow straighteners, and provides an enclosed volume suitable for larger, lower-speed experiments. The third section is a contraction, terminating in a 2" x 2", higher-speed square section. The wind tunnel has a footprint of approximately 13.5" x 5.5", making it small enough to be portable and to fit on a desk. An off-the-shelf masked stereolithography apparatus (MSLA) 3D printer was used to prepare the parts. This allows the wind tunnel to be built for under $500; even including the cost of a 3D printer, the overall cost remains under $1,000. This design is able to produce flow at up to 44.1 m/s, enabling a variety of aerodynamic demonstrations.

## 1. Introduction

Wind tunnels remain an essential tool for aerodynamics investigations; however, constraints associated with cost and overall size can pose a significant practical obstacle to their access. Here we focus on proposing improvements to small wind tunnels, while laying the groundwork for scaling the design up to reduce the overall footprint of larger setups.

A large body of knowledge has been accumulated on wind tunnel design [1][2][3], including aspects such as contraction shape [4], grids and screens for flow uniformity [4][5], and use of numerical simulation to predict performance [6]. There has been recent interest in improving small wind tunnels, with varied applications ranging from animal experiments [7][8] to testing micro-energy harvesters [9]. Notably, [7] introduced a compact design for closed-loop tunnel, by making innovative use of a large number of turning vanes and honeycomb structures to support smooth flow across turns. This design used a single fan. In [9], a blower fan, and a straight-walled diffuser and contraction, separated by a screen, to produce more uniform flow in a blowing, open-section layout. The use of a single large fan or blower can simplify mechanical design, but can also result in significant nonuniformities in the flow leaving the fan, requiring large distances between the fan and the test section.

The main objective of this paper is to present a method for building a cost effective, portable and reliable wind tunnel that can produce optimal conditions for data collection for different scenarios such as in class demonstrations and sensor calibration. An open-return, blowing-type layout was selected to achieve highest possible portability and flexibility of use. To produce more uniform flow from the outset, and therefore reduce the separation distance needed between fan and test section, flow was generated by replacing the conventional single fan with an array of smaller fans. Resin 3D printing allows inexpensively producing complex parts with smooth surface finish. The tunnel design is described in section 2, together with the flow measurement setup. Flow data are reported in section 3. A brief discussion and next steps are outlined in section 4.

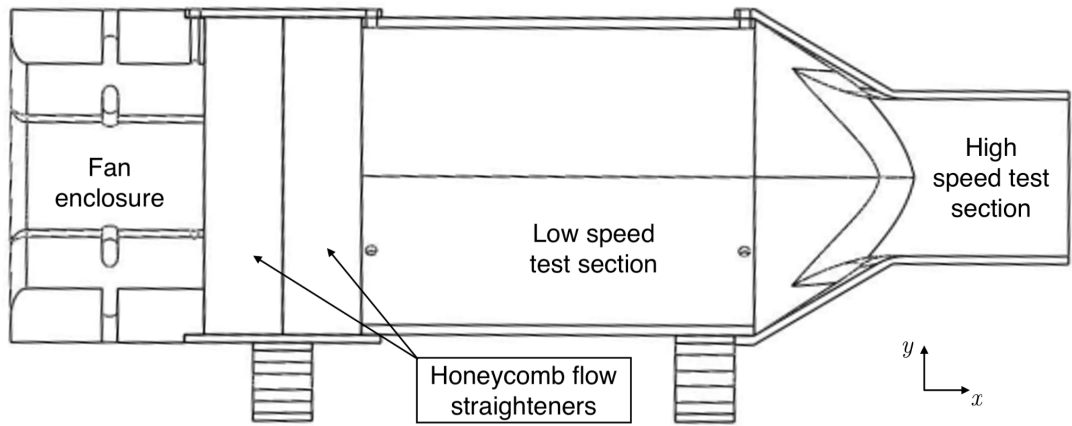

**Figure 1.** Overall wind tunnel schematic.

**2. Materials and Methods**

An open return wind tunnel was chosen because not needing a recirculation system results in a reduced overall size; an overall diagram is provided in Figure 1. The material for the wind tunnel body was a curable resin, processed by an Elegoo Saturn 2 MSLA 3D printer. This produces a layer height of 28.5 micron, making a smooth internal surface and producing gradual transitions in the wind tunnel for optimal flow conditions. This printer has a build volume of 219mm x 123mm x 250mm, which dictates the allowable size of each component. The ducted fans (JFtech QF1611(1311) - 14000KV) are 30 mm in diameter and have 6-bladed propellers (see Figure 2). Each fan has a maximum continuous current of 12A with a maximum thrust of 220g. Each fan was paired with a 30A electronic speed controller (ESC, from RC Electronic Parts) for brushless motors. To vary the speed of the ducted fans, an ESC consistency tester was used. This ESC consistency tester has an output signal width ranging from 800-2200$\mu$s, which controls the speed of the wind tunnel. The electronics are powered by a 1000W AC/DC converter (SE-1000-12, Mean Well USA Inc.) that has a maximum current output of 83.8A.

The fans were arranged in a hexagonal array, which is the arrangement with the closest possible packing (covering 90% of available space), as opposed to a more conventional rectangular array (which would cover 70% of available space). The hexagonal packing was therefore expected to provide a more uniform flow, reducing the need for a longer settling chamber. Butyl rubber was used to manage wiring inside of the fan housing to minimize interference with the flow. The fan housing is designed as a standalone structure that can be removed and altered independently of the rest of the wind tunnel, as shown in figure 2. The diagonal of the hexagon measures 4.4 in (such that one of the sides is 2.2 in long), implying a cross-sectional area of 12.57 in$^2$. (i.e. 81.13 cm$^2$).

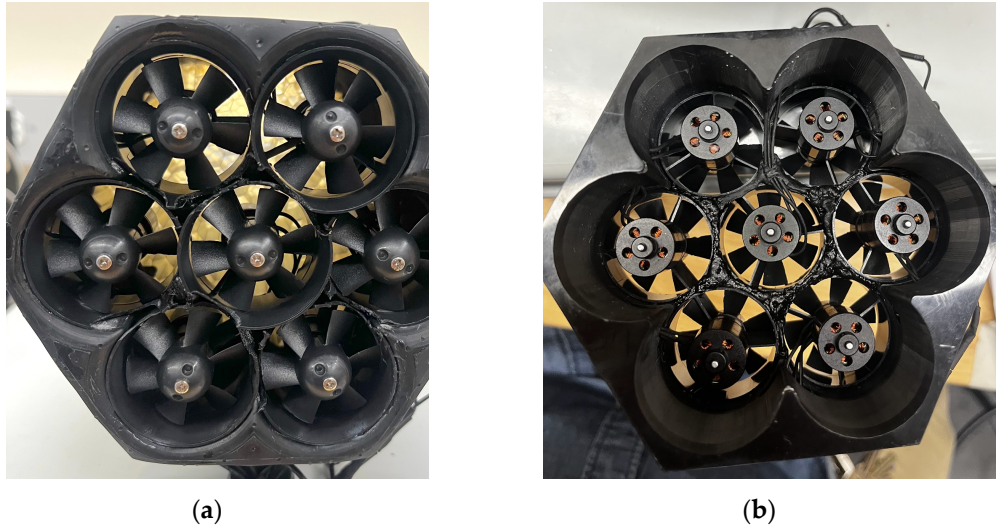

**Figure 2.** Fan Enclosure. **(a)** Fan inlet side. **(b)** Fan outlet side. The enclosure is 3D printed; the ducted fans are held in place with butyl rubber.

The next part of the wind tunnel is the flow conditioning section, which holds two one-inch-thick hexagonal honeycomb flow straighteners, with openings of nominal sizes ¼" and ⅛" respectively, as shown in figures 3(a) and (b). The honeycomb was cut to size using a table saw to ensure clean cuts to prevent blockages from forming at the wind tunnel perimeter (although a hack saw could also be used). Prior work has shown that honeycomb cells provide the best performance for flow straightening if they have a length-to-diameter ratio between 7 and 10 [Nagib?]. This criterion was met by the second honeycomb cell (which had a ratio of approximately 8). The flow conditioning section was also 3D printed as an independent structure, to allow different flow straighteners to be used without a complete redesign of the system. This leads into the settling chamber where low speed tests can also be performed, shown in figure 4(a). Again this was 3D printed independently to allow for later modifications.

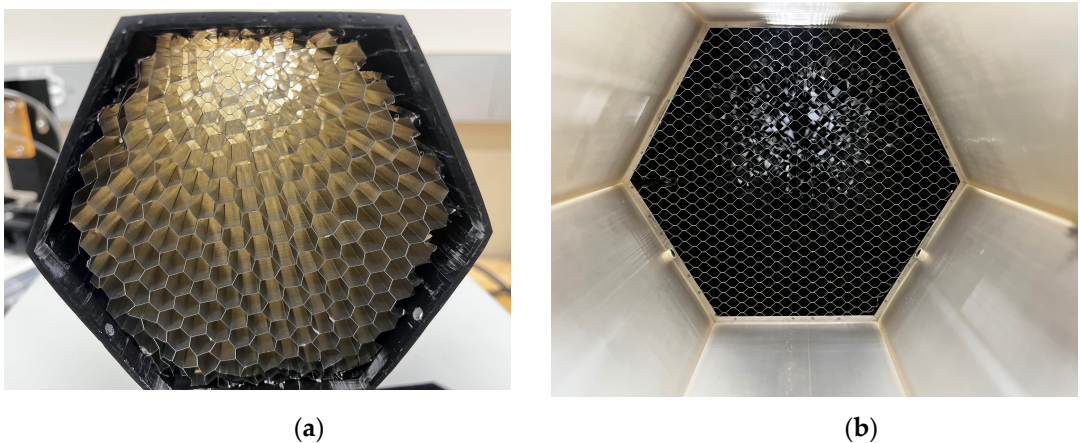

**Figure 3.** Flow Conditioning Section. **(a)** initial honeycomb, **(b)** final honeycomb.

The contraction was printed to match the hexagonal pattern with a smooth transition to a 2" x 2" square high-speed test section (implying a cross-sectional area of 4 in$^2$ = 25.81 cm$^2$), shown in figure 4(b). This again was printed independently to allow for further modifications without affecting the wind tunnel's overall design. The area ratio is 3.14 at the contraction is relatively low, but provides sufficiently uniform flow (as shown below) while keeping an exceptionally small footprint for a given test section size, as discussed in Section 4.

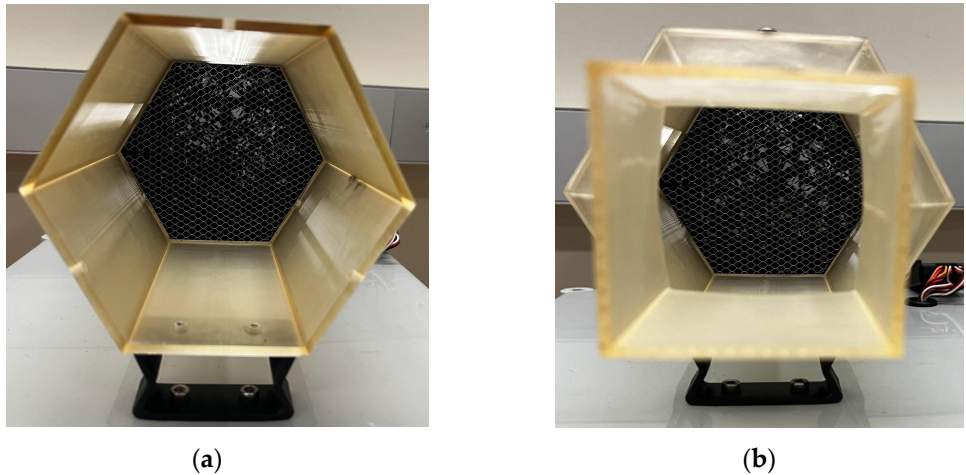

(**a**) (**b**)

**Figure 4. (a)** Low Speed Test Section. **(b)** High Speed Test Section.

Several experiments were performed in this model wind tunnel. First, we tested the uniformity of the flow in the high speed section. This was done using a translating platform that moves incrementally in the $x$ and $z$ directions. A streamlined arm was 3D printed to attach to the platform and to mount a pitot-static tube for velocity measurements. Two differential manometers were used. At lower speeds, a Dwyer 2002 Magnehelic Differential Pressure Gauge provided higher resolution, whereas at higher speeds we switched to a JH Gauge DPGJH20005X. The 2" x 2" high speed test section was split up into nine equal parts and the pitot tube was placed in the center of each part using the translating platform; the measurements were repeated at four speed settings, spanning the range between approximately 15 m/s and 45 m/s. In the next test, we measured flow uniformity in the low speed section of the wind tunnel. For this purpose, the contraction can be removed in seconds by loosening three mounting screws, as shown in figure 4(a). The low speed section was split up into 11 sampling locations because of its hexagonal shape.

### 3. Results

#### 3.1. High speed section

As described above, air velocity data was obtained using a pitot-static probe. Once this data was collected, a percent difference calculation was performed to determine the deviation from the average flow velocity at each point. Results can be seen in figure 6. The greatest difference in the low speed data was observed in the center, where readings are up to a maximum of 3.4% below the average. One possible solution for this would be to add independent controllers to each motor. This would allow the user to vary the output until perfectly uniform flow is recorded. This would add complexity to the system and introduce an additional element requiring calibration, and was therefore not pursued here.

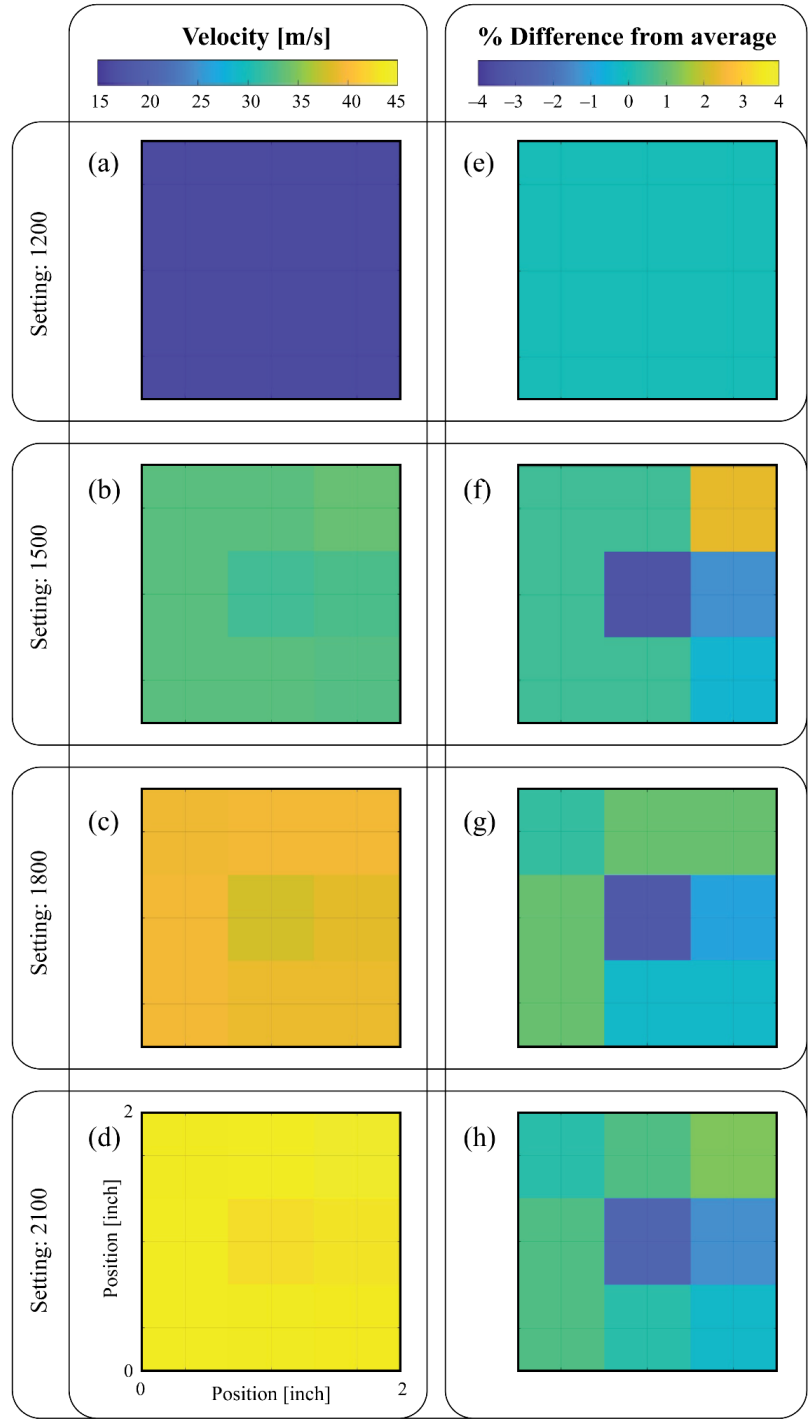

**Figure 6.** Flow uniformity in the high speed section, determined by measuring velocity **(a–d)** and calculating percent difference from the average value **(e–f).**

*3.2. Low speed section*

For completeness, mean flow measurements were also taken in the low speed selection. The resulting data can be seen in figure 7. The velocity here is much less uniform when compared to the high speed section. When analyzing the percent difference from the average velocity the maximum observed difference is about 45 percent.

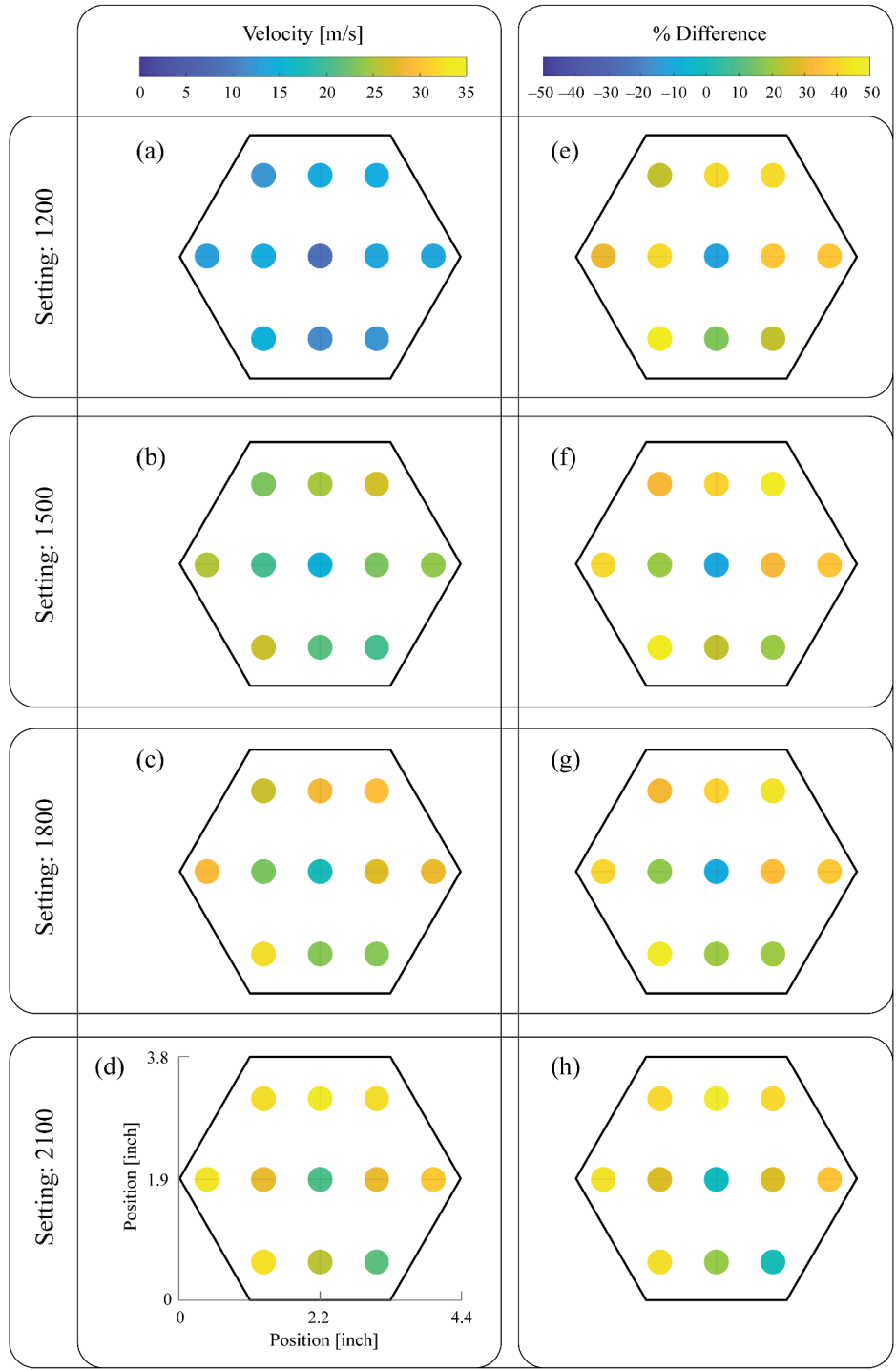

**Figure 7.** Flow uniformity in the low-speed section, determined by measuring velocity **(a–d)** and calculating percent difference from the average value **(e–f)**.

## 4. Discussion

A portable wind tunnel, with a footprint of 13.5" x 5.5", was built and tested. The wind tunnel used an array of seven hexagonally-packed ducted fans (as opposed to a single fan, as for conventional designs) to generate a relatively uniform flow. A moderate contraction area ratio of 3.14 was used. Flow in the test section exhibited a maximum spatial variation of 3.4%, measured over speeds ranging between 15 m/s and 45 m/s. The tunnel was fabricated using an SLA 3D printer, and ducted fans and electronic speed controllers originally intended for RC aircraft were used. This design and construction approach also allowed keeping the overall cost under $1,000. The resulting tunnel can be used for small-scale experiments, instrument calibration, as well as instruction and rapid even in-class demonstrations.

Next steps include developing a procedure to differentially adjust fans to induce more uniform flow, as well as to generate prescribed velocity profiles – such as thick boundary layers. This wind tunnel design can also be scaled up to construct wind and water tunnels with large test sections, which nevertheless have a much more compact overall footprint than conventional designs.